\documentclass[12pt,preprint,natbib,iop,numberedappendix,appendixfloats]{emulateapj}
\usepackage{graphicx,epsfig,amsmath}

\lefthead{van der Wel et al.}
\righthead{The LEGA-C Survey}
\slugcomment{Accepted for publication in ApJ}
\begin{document}

\newcommand{\gala}{{\tt GALAPAGOS}~}
\newcommand{\gf}{{\tt GALFIT}~}
\newcommand{\mh}{H_{\rm{F160W}}}
\newcommand{\kms}{\>{\rm km}\,{\rm s}^{-1}}
\newcommand{\reff}{R_{\rm{eff}}}
\newcommand{\msol}{M_{\odot}}
\newcommand{\msola}{10^{11}~M_{\odot}}
\newcommand{\msolb}{10^{10}~M_{\odot}}
\newcommand{\msolc}{10^{9}~M_{\odot}}
\newcommand{\ang}{\rm{\AA}}

\title{The VLT LEGA-C Spectroscopic Survey: The Physics of Galaxies at a Lookback Time of 7 Gyr}

\author{A.~van der Wel\altaffilmark{1}}
\author{K.~Noeske\altaffilmark{1}}
\author{R.~Bezanson\altaffilmark{2}}
\author{C.~Pacifici\altaffilmark{3}}
\author{A.~Gallazzi\altaffilmark{4}}
\author{M.~Franx\altaffilmark{5}}
\author{J.C.~Mu\~noz-Mateos\altaffilmark{6}}
\author{E.F.~Bell\altaffilmark{7}}
\author{G.~Brammer\altaffilmark{8}}
\author{S.~Charlot\altaffilmark{9}}
\author{P.~Chauk\'e\altaffilmark{1}}
\author{I.~Labb\'e\altaffilmark{5}}
\author{M.V.~Maseda\altaffilmark{5}}
\author{A.~Muzzin\altaffilmark{10}}
\author{H.-W.~Rix\altaffilmark{1}}
\author{D.~Sobral\altaffilmark{11},\altaffilmark{5}}
\author{J.~van de Sande\altaffilmark{12}}
\author{P.G.~van Dokkum\altaffilmark{13}}
\author{V.~Wild\altaffilmark{14}}
\author{C.~Wolf\altaffilmark{15}}

\altaffiltext{1}{Max-Planck Institut f\"ur Astronomie, K\"onigstuhl
  17, D-69117, Heidelberg, Germany; e-mail:vdwel@mpia.de}

\altaffiltext{2}{Steward Observatory, University of Arizona, 933 North
  Cherry Avenue, Tucson, AZ 85721, USA}

\altaffiltext{3}{Astrophysics Science Division, Goddard Space Flight
  Center, Code 665, Greenbelt, MD 20771, USA}

\altaffiltext{4}{INAF-Osservatorio Astrofisico di Arcetri, Largo
  Enrico Fermi 5, I-50125 Firenze, Italy}

\altaffiltext{5}{Leiden Observatory, Leiden University, P.O.Box 9513,
  NL-2300 AA Leiden, Netherlands}

\altaffiltext{6}{European Southern Observatory, Alonso de Córdova
  3107, Casilla 19001, Vitacura, Santiago, Chile}

\altaffiltext{7}{Department of Astronomy, University of Michigan, 1085
  S.~University Ave, Ann Arbor, MI 48109, USA}

\altaffiltext{8}{Space Telescope Science Institute, 3700 San Martin
  Drive, Baltimore, MD 21218, USA}

\altaffiltext{9}{UPMC-CNRS, UMR 7095, Institut d'Astrophysique de
  Paris, F-75014 Paris, France}

\altaffiltext{10}{Institute of Astronomy, University of Cambridge,
  Madingley Road, Cambridge, CB3 0HA, UK}

\altaffiltext{11}{Department of Physics, Lancaster University,
  Lancaster, LA1 4YB, UK}

\altaffiltext{12}{Sydney Institute for Astronomy, School of Physics,
  University of Sydney, NSW 2006, Australia}

\altaffiltext{13}{Department of Astronomy, Yale University, New Haven,
  CT 06511, USA}

\altaffiltext{14}{School of Physics and Astronomy, University of St
  Andrews, North Haugh, St Andrews, KY16 9SS, UK}

\altaffiltext{15}{Research School of Astronomy and Astrophysics,
  Australian National University, Canberra, ACT 2611, Australia}
  
\begin{abstract}

The Large Early Galaxy Census
(LEGA-C\footnote{http://www.mpia.de/home/legac/index.html}) is a
Public Spectroscopic Survey of $\sim3200$ $K$-band selected galaxies
at redshifts $z=0.6-1.0$ with stellar masses $M_*>\msolb$, conducted
with VIMOS on ESO's Very Large Telescope. The survey is embedded in
the COSMOS field ($R.A. = 10h00$; $Dec.=+2\deg$).  The 20-hour long
integrations produce high-$S/N$ continuum spectra that reveal ages,
metallicities and velocity dispersions of the stellar populations.
LEGA-C's unique combination of sample size and depth will enable us
for the first time to map the stellar content at large look-back time,
across galaxies of different types and star-formation activity.
Observations started in December 2014 and are planned to be completed
by mid 2018, with early data releases of the spectra and value-added
products.  In this paper we present the science case, the observing
strategy, an overview of the data reduction process and data products,
and a first look at the relationship between galaxy structure and
spectral properties, as it existed 7 Gyr ago.
\end{abstract}

\section{Introduction}

Our understanding of galaxy evolution at late cosmic epochs has been
revolutionized by spectra from the Sloan Digital Sky Survey
\citep{york00}: their high signal-to-noise and high resolution have
allowed accurate measurements of the (light-weighted) integrated
stellar ages and metallicities of galaxies
\citep[e.g.,][]{kauffmann03, gallazzi05}.  These data have brought
into clear focus the multi-variate correlations between stellar
population properties and mass \citep{kauffmann03a}, structure
\citep[e.g., concentration, bulge-to-disk ratio;][]{kauffmann03a},
size \citep{van-der-wel09}, stellar velocity dispersion
\citep{gallazzi06,graves09a}, nuclear activity \citep{kauffmann03b,
  schawinski09}, and environment \citep{kauffmann04, pasquali10,
  thomas10}, which have greatly illuminated the processes that drive
star-formation and ongoing assembly of present-day galaxies.  The main
limitation of examining present-day galaxies for the purpose of
reconstructing their formation history is that most of the star
formation occurred in the distant past: mean stellar ages of $L^*$
galaxies are typically well over 5~Gyr and it is difficult to resolve
star formation histories from integrated spectra.

Now LEGA-C is obtaining spectra of similar quality for $\sim$3200
$K$-band selected galaxies in the redshift range $0.6<z<1.0$, at a
look-back time of $6-8$ Gyr.  LEGA-C is a 4-year, 128-night Public
Spectroscopic Survey with VLT/VIMOS \citep{le-fevre03}, one of the
most extensive extra-galactic surveys on an 8-10m-class telescope, and
unique in its combination of depth and resolution. We know that
$\sim$50\% of all stars in the universe formed since $z\sim 1$
\citep[e.g.,][]{dickinson03, rudnick03, ilbert10, muzzin13}.  As a
consequence, 8 Gyr ago galaxies must have had very different stellar
populations. Yet, we have not diagnosed spectroscopically those
properties for samples of more than a few dozen individual objects
\citep{jorgensen05, jorgensen13, choi14, gallazzi14, belli15}, and we
do not know what processes drive the star-formation and assembly
history of the galaxy population.  The LEGA-C spectra provide us with
a unique opportunity to combine the advantages of the `archaeological'
approach to galaxy formation with the power of the look-back approach.

We know that galaxies at $z\sim 1$ (and beyond) show a wide variety in
structure and star-formation activity \citep[e.g.,][]{conselice11a,
  mortlock13}, but also that, like in the present-day universe, these
properties are correlated \citep[e.g.,][]{trujillo06, franx08, bell12,
  van-der-wel14}.  Knowledge of the stellar ages and metallicities is
crucial to understand the star formation history, and stellar velocity
dispersions are a crucial element in quantifying the dynamical scaling
relations such the \citet{faber76} relation, the \citet{tully77}, and
the fundamental plane \citep{djorgovski87, dressler87}, the evolution
of which have proved to be key in constraining galaxy formation models
\citep[e.g.,][]{franx93, van-dokkum98, holden05, van-der-wel05,
  treu05, bezanson13, bezanson15}.  Furthermore, the underlying
relationships between the stellar content of galaxies and their mass,
ongoing star formation, internal structure, environment, and nuclear
activity can be explored at large look-back time for the first time by
the LEGA-C dataset.

The LEGA-C survey can play a crucial and unique role in the context of
previous, current, and upcoming developments in the field of galaxy
evolution.  Wide-area surveys with HST such as COSMOS
\citep{scoville07a} and CANDELS \citep{grogin11,koekemoer11} have been
completed, producing a high-resolution imaging data set, essentially
`frozen' until JWST and Euclid start operation.  After several years
with a focus on deep imaging campaigns, the deep spectroscopic effort
to probe the stellar light of distant galaxies has fallen behind.
Furthermore, ALMA is measuring cold gas masses for increasingly large
samples of galaxies \citep[e.g.,][]{carilli13, decarli14, scoville15}.
The interplay between stellar populations, ongoing star formation and
the available reservoirs for future star formation make for a powerful
combination to reconstruct and predict galaxy evolution.  Finally, in
2019 the JWST should start operation, marking the beginning of a new
era in which JWST and new, large ground-based telescopes will explore
the physical properties of galaxies at $z\sim 2$ and beyond.  Because
the vast majority of all stars formed between $z\sim 2$ and the
present, connecting the galaxy populations at $z>2$ and the present is
an intractable problem without the intermediate redshift benchmark
sample that LEGA-C provides.  Yet, JWST is lacking spectroscopic
capabilities at $\lambda<1\mu$m, leaving many established spectral
diagnostics of stellar populations inaccessible; crucially LEGA-C
fills this gap.

The main goal of this paper is to introduce the LEGA-C survey. In
particular, we outline its scientific goals in Section
\ref{sec:goals}), describe in detail the survey design, observational
strategy and data reduction/analysis in Section \ref{sec:data}, and
present first-look results in Section \ref{sec:firstlook}.

\section{Goals of LEGA-C}\label{sec:goals}

The LEGA-C observations produce $S/N>10~\rm{\AA}^{-1}$ spectra for
$\sim$3200 $K$-band selected galaxies with redshifts in the range
$0.6<z<1.0$ (or a look-back time of $6-8$ Gyr) and over the wavelength
range $\sim6300{\rm{\AA}}-8800{\rm{\AA}}$.  Our simple $K$-band
selection (see below) is close to a stellar mass selection and blind
to almost any other galaxy property.  We chose a $K$-band limit
instead of a (model-dependent) stellar mass limit to conserve the
legacy value of the dataset even if our current methods to estimate
stellar masses were to prove incomplete or systematically incorrect.
The LEGA-C spectra provide accurate measurements of
\begin{itemize}
\item{Balmer absorption and emission line strengths;}
\item{Metal absorption and emission features;}
\item{Stellar and ionized gas velocity dispersions;}
\item{Emission line fluxes, widths and ratios.}
\end{itemize}
These measurements have a wide range of applications that are of
interest to a large segment of the community, and each of the topics
described below constitutes a large science program. Here, we only
list the main features of each science case.

\subsection{Stellar Population Ages}

The strength of the Balmer absorption lines and the spectral region
around D4000 are the most sensitive diagnostics of stellar population
age \citep[e.g.,][]{kauffmann03, wild09}.  The age distribution of
stellar populations at a look-back time of $6-8$ Gyr is one of the main
legacy data products of the survey.  \citet{gallazzi05} and
\citet{thomas10} measured the age distribution for present-day
galaxies.  Combined with the observed distribution of star-formation
across the mass function \citep[e.g.,][]{karim11} this can be used to
predict the evolution of the age distribution under different
assumptions (growth through \textit{in situ} star formation only;
merging).  By measuring the distribution of stellar ages 7 Gyr ago
\citep[e.g.,][]{gallazzi14} the creation of stellar mass and its
migration across the mass function can be reconstructed.  In
particular, this allows us to address the following, inter-related
topics.

\subsubsection{Quenching}

The continuous increase in the number density of passive galaxies from
$z=1$ to the present \citep[e.g.,][]{bell04a, faber07} implies that
significant numbers of star-forming galaxies have their star-forming
activity quenched and then largely remain quiescent.  What causes this
and how quickly this happens, and whether it is preceded by a burst of
star formation is widely discussed, but remains an unsettled problem.

AGN feedback, owing to large available energies and evidence for AGN
driven outflows and bubbles \citep[e.g.,][]{mcnamara07, nesvadba08,
  genzel14} has long been a strong candidate for driving Galaxy wide
quenching of star formation \citep[e.g.,][]{di-matteo05, croton06,
  hopkins06}.  This idea is consistent with the correlation of
quiescence with the amount of stellar mass in galaxy centers:
concentration \citep{kauffmann03a, franx08, bell08, bell12}, velocity
dispersion \citep{franx08, wake12}, central stellar mass surface
density \citep{fang13}, or bulge mass \citep{lang14}; in the local
universe all these quantities correlate strongly with black hole mass,
lending support to the AGN feedback picture.  However, a direct causal
connection has been difficult to prove, and perhaps the mere presence
of a dense body of evolved stars prevents the formation of a large
number of new stars \citep[e.g.,][]{martig09, conroy15}.

The demographics of recently quenched galaxies shed light on this: the
fraction of galaxies displaying spectral signs of rapidly reduced
recent star formation activity and preceding bursts (e.g., strong
Balmer absorption lines, but no emission lines -- e.g., \citet{wild09}
-- constrains the quenching rate, speed, and mode.

\subsubsection{Star Formation and Galaxy Structure}

The connection between quiescence and galaxy structure is a specific
instance of a general connection between star formation history and
galaxy structure: among star-forming galaxies the instantaneous SFR
depends on concentration \citep{whitaker15}, and it has been argued
that bulge fraction is the driving factor \citep{abramson14, lang14}.
This explains to some extent, but not fully, the scatter in SFR at
fixed galaxy mass.  With age measurements for star-forming galaxies
the contribution of temporal variations in SFR to the scatter in the
so-called star-forming Main Sequence \citep{noeske07} can be
constrained.  Furthermore, as ALMA will start measuring cold gas
masses of large samples of galaxies \citep{decarli14, scoville15} the
correlations between star formation, galactic structure and gas
fraction provide additional constraints on the short- and long-term
variations in star formation activity.

\subsection{Metallicities}

The LEGA-C dataset provides the first simultaneous systematic
measurement of the stellar and gas phase metal content of the galaxy
population at a large look-back time.

\subsubsection{Stellar metallicity}

The strength of stellar absorption features such as the Fe and Mg
lines constrain the metal content and, hence, the metal enrichment
history of galaxies as they are produced at different stages of the
evolution of a stellar population.  Through the comparison with the
present-day galaxy population, and in combination with the age
measurements, the abundance measurements put new and independent
constraints on the amount of star formation and chemical enrichment of
galaxies over the past 8 Gyr.  Previous studies at $z>0.5$ were
limited to small samples of very high-mass galaxies or used co-added
spectra \citep{schiavon06, jorgensen13, gallazzi14}.  With the large
dynamic range in mass and larger sample size we can now measure the
intrinsic scatter and mass dependence, which is crucial for
interpreting the redshift evolution of the mass-metallicity relation.
For several hundred spectra with $S/N>30 \rm{\AA}^{-1}$ the
$\alpha$/Fe abundance ratio can be quantified.  This reveals to what
extent the most massive galaxies have had their stellar populations
diluted by late star formation and accretion of externally formed
stars (mergers).

\subsubsection{Gas phase metallicity}

The standard indicators of the oxygen abundance (as a proxy of
gas-phase metallicity) H$\beta$, $[$OII$]\lambda3727$, and
$[$OIII$]\lambda5007$ \citep{kewley08} are available for the
$\sim$1300 galaxies at redshift $z\sim 0.8$ and below.  For the
remainder one must rely on the other Balmer lines and
$[$OII$]\lambda3727$.  The detection of (or useful upper limit on) the
faint $[$OIII$]\lambda4363$ line in the deep spectra constrains the
electron temperature and therefore breaks the degeneracy between
metallicity and electron temperature.

\subsection{Kinematics}

\subsubsection{Dynamical masses}

Our knowledge of the evolution of the mass function of galaxies relies
on photometry-based stellar mass estimates, which depend on a range of
assumptions regarding the star-formation history, the metal content,
the dust content (and its extinction law and geometry), and the
stellar initial mass function.  Comparisons with dynamical mass
measurements have been carried out across galaxies of different types
in the present-day universe \citep[e.g.,][]{gallazzi06,taylor10a}, but
the necessary measurements (stellar velocity dispersions) have so far
been lacking at higher redshifts, with the exception of small samples
of massive early-type galaxies \citep{van-der-wel06a, rettura06,
  kannappan07, van-de-sande15}.

The 5-10\% accurate stellar velocity dispersions from the high-$S/N$,
high-resolution ($\sim50$km$/$s) LEGA-C spectra, combined with
HST-based sizes and Sersic indices, provide precise dynamical mass
measurements.  The dynamical masses calibrate stellar mass
measurements.  Furthermore, differences between dynamical and stellar
mass estimates can constrain contributions from dark matter, gas, and
IMF variations, and be distinguished by means of cold gas measurements
from ALMA, increasing further the accuracy of the work based on gas
dynamics \citep[e.g.,][]{erb06, forster-schreiber09, law09, epinat09,
  erb14}.

\subsubsection{Scaling relations}

The measurement of stellar velocity dispersions and the already
available structural parameters from HST imaging allow a detailed
description of the dynamical scaling relations.  The \citet{faber76}
relation and Fundamental Plane \citep{dressler87, djorgovski87} for
early-type galaxies are mapped out at $z\sim 1$ over a large range in
galaxy mass of over an order in magnitude, addressing the lingering
dispute on the evolution of the Fundamental Plane tilt
\citep{van-der-wel05, treu05, holden10}.

For late-type galaxies the \citet{tully77} relation can be
reconstructed by statistically deprojecting the emission and/or
absorption line widths to derive the circular velocity \citep{rix97a,
  kassin07, maseda13}.  The LEGA-C dataset provides the important
opportunity to test ionized gas kinematics with stellar kinematics and
calibrate dynamical masses based on emission line widths, building
further on previous work \citep[e.g.,][]{vogt96, ziegler02, bohm07,
  kassin07, van-starkenburg08, puech08, gnerucci11, miller11,
  miller12}.  The evolution of these scaling relations across 8 Gyr
provides stringent constraints on the growth modes of galaxies such as
inside-out disk growth \citep{miller11, dutton11} and merging
\citep{naab09a, johansson12}.

\begin{figure*}
  \includegraphics[scale=.43]{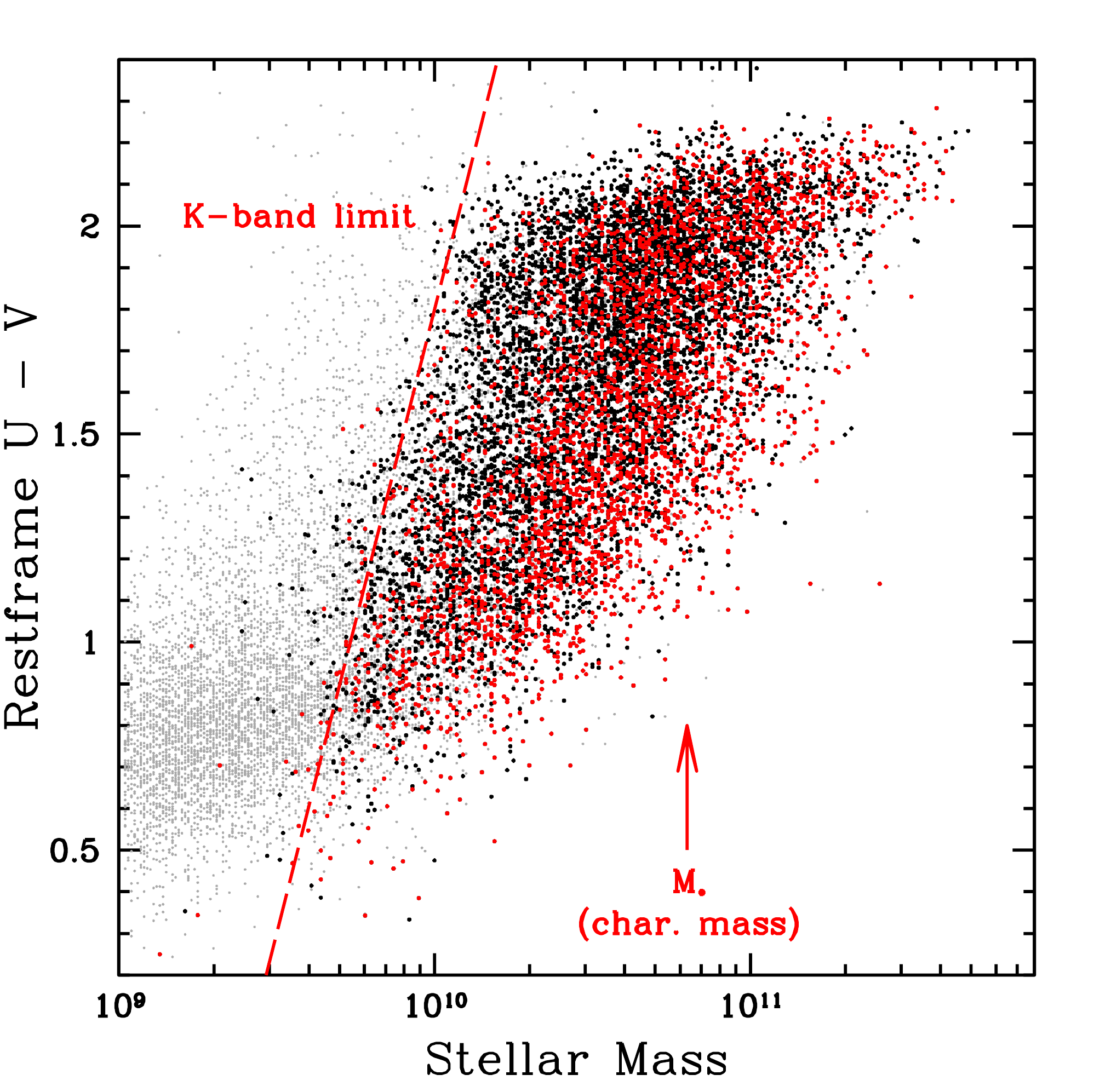}
  \includegraphics[scale=.43]{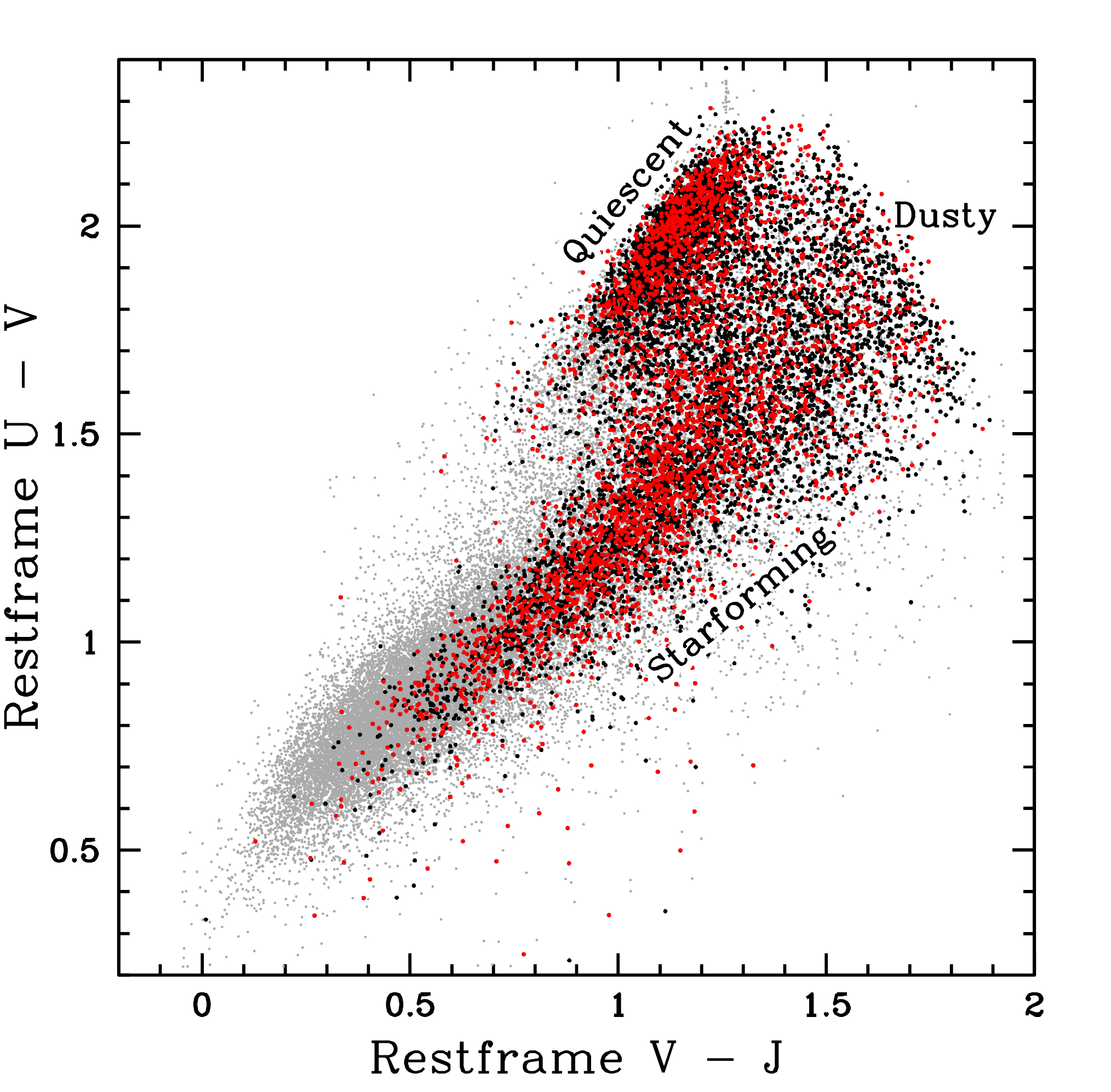}
  \caption{Stellar mass and color distribution of the galaxies in the
    LEGA-C primary sample.  In light gray we show the full UltraVISTA
    catalog from \citep{muzzin13a} for the redshift range $0.6<z<1.0$.
    In black we show the $K$ band-limited parent sample -- the red
    dashed line in the left-hand panel shows the approximate mass
    limit that corresponds with our $K$-band limit -- and in red we
    overlay the primary sample that is included in the survey.  The
    sample straddles the characteristic galaxy mass, $M^*$, and spans
    the full observed color range, that is, SED type.  }
\label{fig:sample}
\end{figure*}

\section{Data}\label{sec:data}
\subsection{Sample Selection}\label{sec:sample}

The parent sample is the photometric sample from \citet{muzzin13a}.
Their UltraVISTA $K$-band selected catalog is 90\% complete down to
$K=23.4$ and contains 160,070 sources across 1.62 square degrees in
the COSMOS field (R.A.$=10^{\rm h} 00^{\rm m} $; Dec.$=+2^{\circ}
12\arcmin$).  On top of the near-IR VISTA imaging data
\citep{mccracken12}, Muzzin et al.~collected and consolidated the
available photometric information across 30 passbands ranging over 2
orders of magnitude in wavelength, from 0.2$\mu$m to 24$\mu$m
micron. Photometric redshifts were measured by Muzzin et al.~using
EAZY \citep{brammer08}.  Cross-matching with the \citet{davies15}
catalog results in 21,267 galaxies with spectroscopic redshifts.

\begin{table*}
\begin{center}
  \caption{ Sample selection }
 \begin{tabular}{|l|c|c|c|c|}
\hline
Sample & Redshift & $K$-band limit & \#~in UltraVISTA & \#~in LEGA-C \\
\hline
Parent         &     \nodata     & 23.4 (90\% compl.)  &      160,070      &   4,027     \\
Primary        &     $0.6<z<1.0$ & $< 20.7-7.5\log((1+z)/1.8)$  &      9,920   &       3,142 \\
Filler I       &     $1.0<z<1.5$ & $< 20.4$                     &      2,224   &       159 \\
Filler II      &     $0.6<z<1.0$ & $> 20.7-7.5\log((1+z)/1.8)$  &      36,499  &       630 \\
Filler III     &     $1.0<z<1.5$ & $> 20.4$                     &      37,136  &       96 \\
\hline
\end{tabular}
\label{tab:sample}
\end{center}
\end{table*}

Our sample selection procedure is summarized in Table 1.  Our primary
sample\footnote{The primary and filler samples are pre-selected to
  include only those objects in the UltraVISTA catalog that are not
  classified as stars and have USE flag with value 1, indicating `good
  photometric quality' -- see \citet{muzzin13a} for details.} for the
selection of LEGA-C targets consists of 9,920 galaxies brighter than
$K =20.7-7.5\times\log((1+z)/1.8)$ and $0.6 < z < 1.0$ (spectroscopic
or photometric).  That is, our primary sample is K-band selected, with
a redshift-dependent magnitude limit that ranges from $K=21.08$ at
$z=0.6$ to $K=20.7$ at $z=0.8$ to $K=20.36$ at $z=1.0$.  The
redshift-dependent K-band limit ensures that primary targets are
sufficiently bright in the spectroscopically observed wavelength range
($0.6\mu m - 0.9\mu m$) to produce a high-quality spectrum, while
retaining the advantages of selecting galaxies in the rest-frame
near-IR, namely the reduced dependence on variations in age,
star-formation activity and extinction.

The basic physical parameters of the galaxies in the primary sample
are shown in Figure \ref{fig:sample}.  For all stellar masses larger
than $\sim\msolb$ we sample the full range of SED types, from blue to
dusty, from star-forming to passive.  We note that thanks to our
$K$-band selection (rather than, e.g., stellar mass) our sample is not
strongly biased against particular sub-types of galaxies, such as
extremely obscured or emission line-dominated objects.

The sample size of $\sim3200$ and the mass range are motivated by the
measurement precision and by the goal of studying the intrinsic
scatter in galaxy properties.  We require these to be the same: in
order to decide whether two sets of galaxies differ in terms of, e.g.,
age by more than the intrinsic scatter in age, each set needs to
contain $\sim$80 galaxies (for a 3$\sigma$ result).  The galaxies span
a range in stellar mass of over 1.5 orders of magnitude (Figure
\ref{fig:sample}, \textit{left}).  Given measurement uncertainties of
0.1-0.2 dex in stellar and dynamical masses, sets of galaxies should
be chosen within $\sim0.3$~dex mass ranges. That is, the survey will
probe 6 independent mass bins.  The galaxy population also shows a
complex structure in color-color space (Figure \ref{fig:sample},
\textit{right}). Given $\sim$0.10-0.15 mag uncertainties in the
colors, sets of galaxies should be chosen within $\sim$0.3 mag-wide
color ranges.  Details of the exact definition of color-color bins
aside, approximately 6 independent color-color bins are needed to
sample the observed range of SED types.  The 6 mass and color-color
bins make a total of 36 bins, which each should contain $\sim$80
galaxies.  This motivates our total sample size of about three
thousand galaxies.

\begin{figure*}
  \includegraphics[scale=.43]{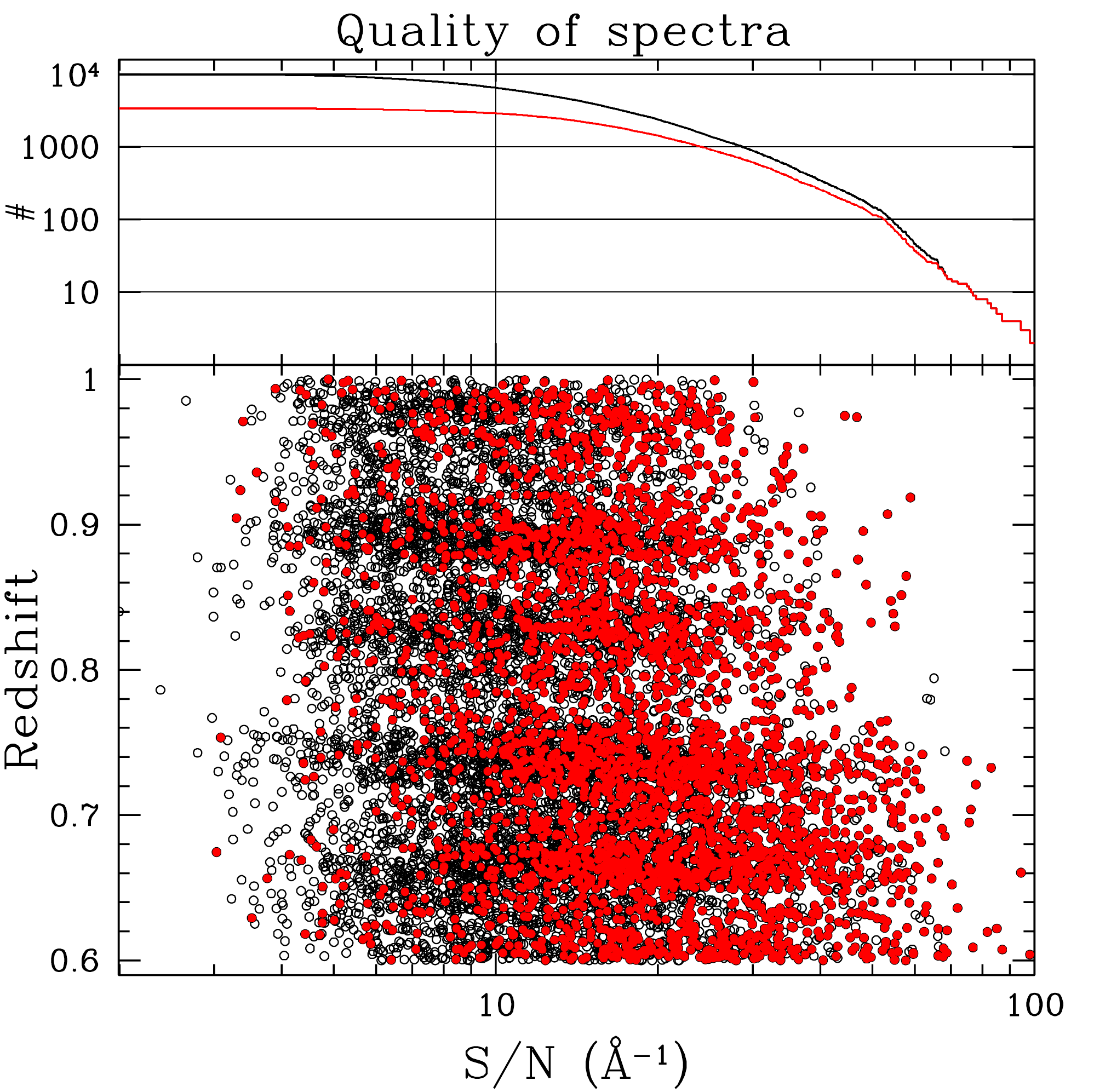}
  \includegraphics[scale=.43]{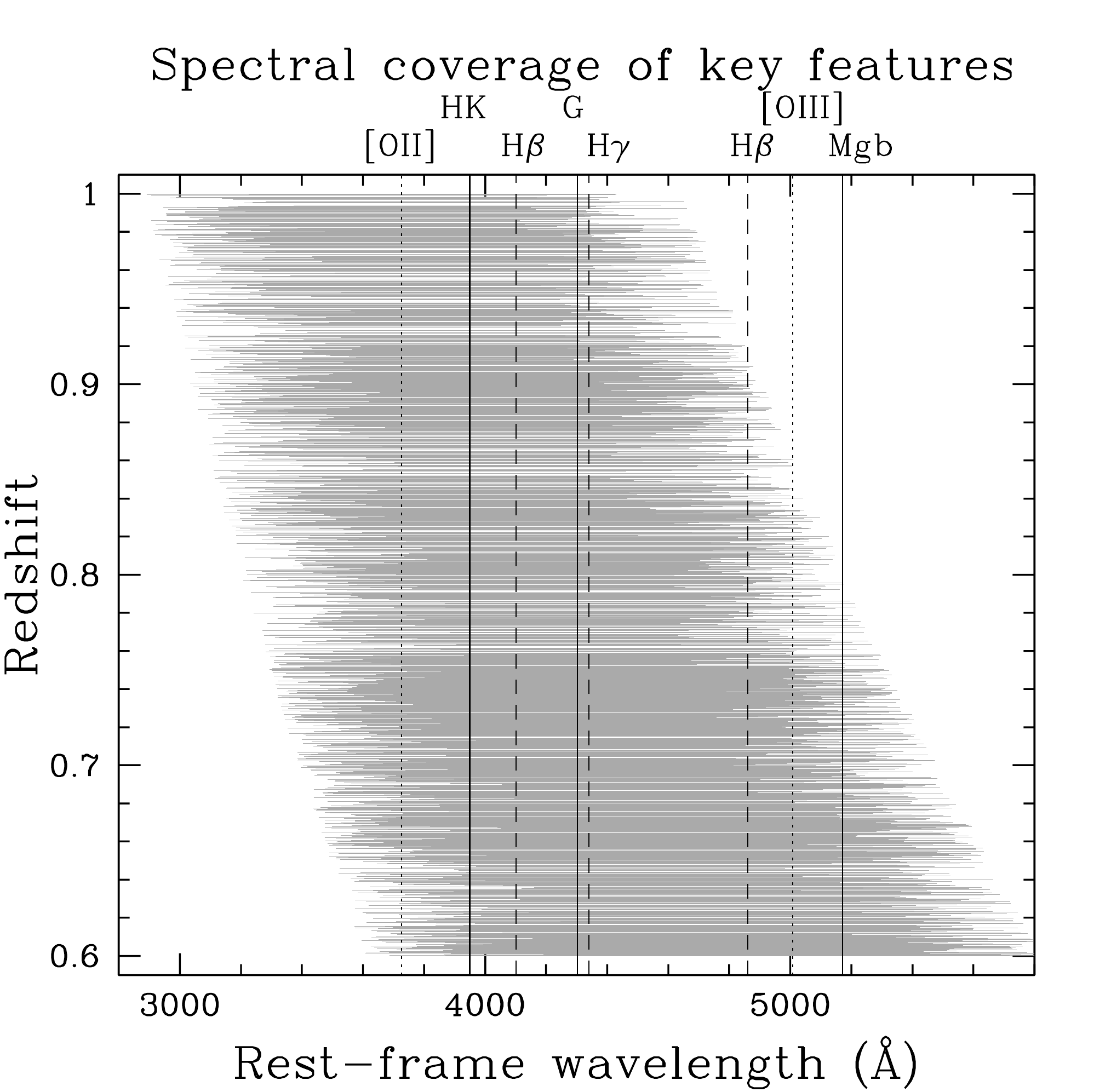}
\caption{\textit{Left:} Distribution of expected $S/N$ vs.~redshift
  for our primary sample.  The points represent the $K$ band- selected
  parent UltraVISTA sample; the red points are those that are in the
  LEGA-C sample.  The top, inset panel shows the cumulative number of
  galaxies as a function of minimum $S/N$, black for the parent
  sample, red for the observed sample.  \textit{Right:} Rest-frame
  wavelength range for the spectra of the observed galaxies. The
  variation at fixed redshift is due to the variation in slit
  position.  Vertical lines indicate the location of important
  spectral features.}
\label{fig:quality}
\end{figure*}

The required depth of the data, obtained with 20h of integration time,
is motivated by the required precision of the measurements, which in
turn is dictated by the intrinsic variation of the physical
parameters.  For age and metallicity the intrinsic scatter for
present-day galaxies is 0.2 dex or more \citep[depending on
  mass,][]{gallazzi05}.  That is, we need measurement uncertainties of
0.2 dex or less in these parameters.  Similarly, the intrinsic scatter
in velocity dispersion for galaxies of a given mass is 0.1 dex
\citep[e.g.,][]{van-der-wel08b}, demanding measurement errors less
than that.  Our data in hand confirm that the obtained signal-to-noise
ratio is as high as anticipated (Section \ref{sec:firstlook}) and
sufficient for velocity dispersion, age, and metallicity measurements
with a precision of 0.08, 0.1, and 0.2 dex, respectively.

Additional samples are used to fill remaining mask real estate.  The
filler sample I consists of galaxies at $1.0 < z < 1.5$ brighter than
$K=20.4$; the filler sample II consists of galaxies at $0.6 < z < 1.0$
but fainter than the (redshift-dependent) K-band limit applied to the
primary sample. Any remaining mask space is used for galaxies in the
redshift range $1.0 < z < 1.5$ and fainter than $K=20.4$ (filler
sample III).

\subsection{Mask Design}

\begin{figure*}
  \includegraphics[scale=.43]{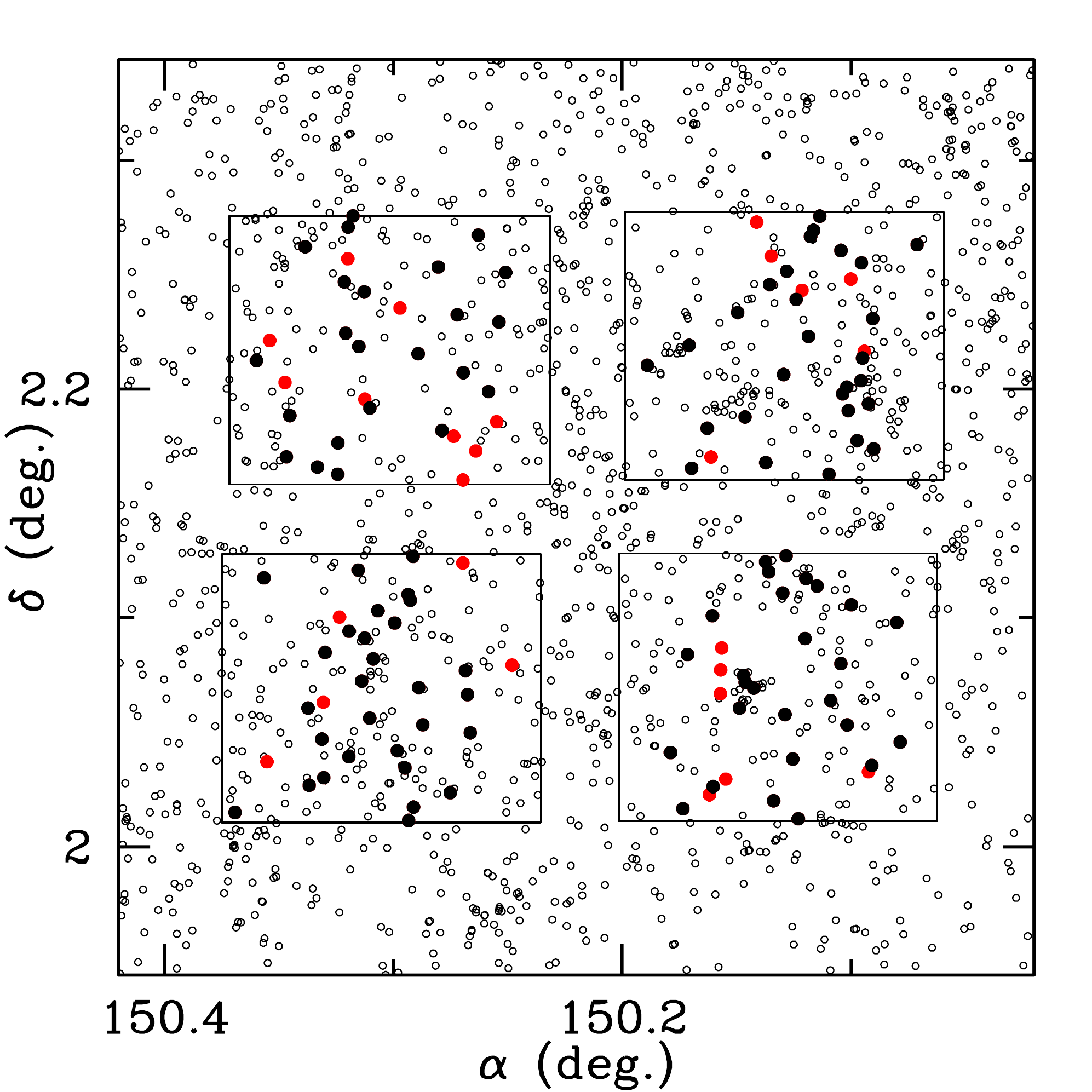}
  \includegraphics[scale=.43]{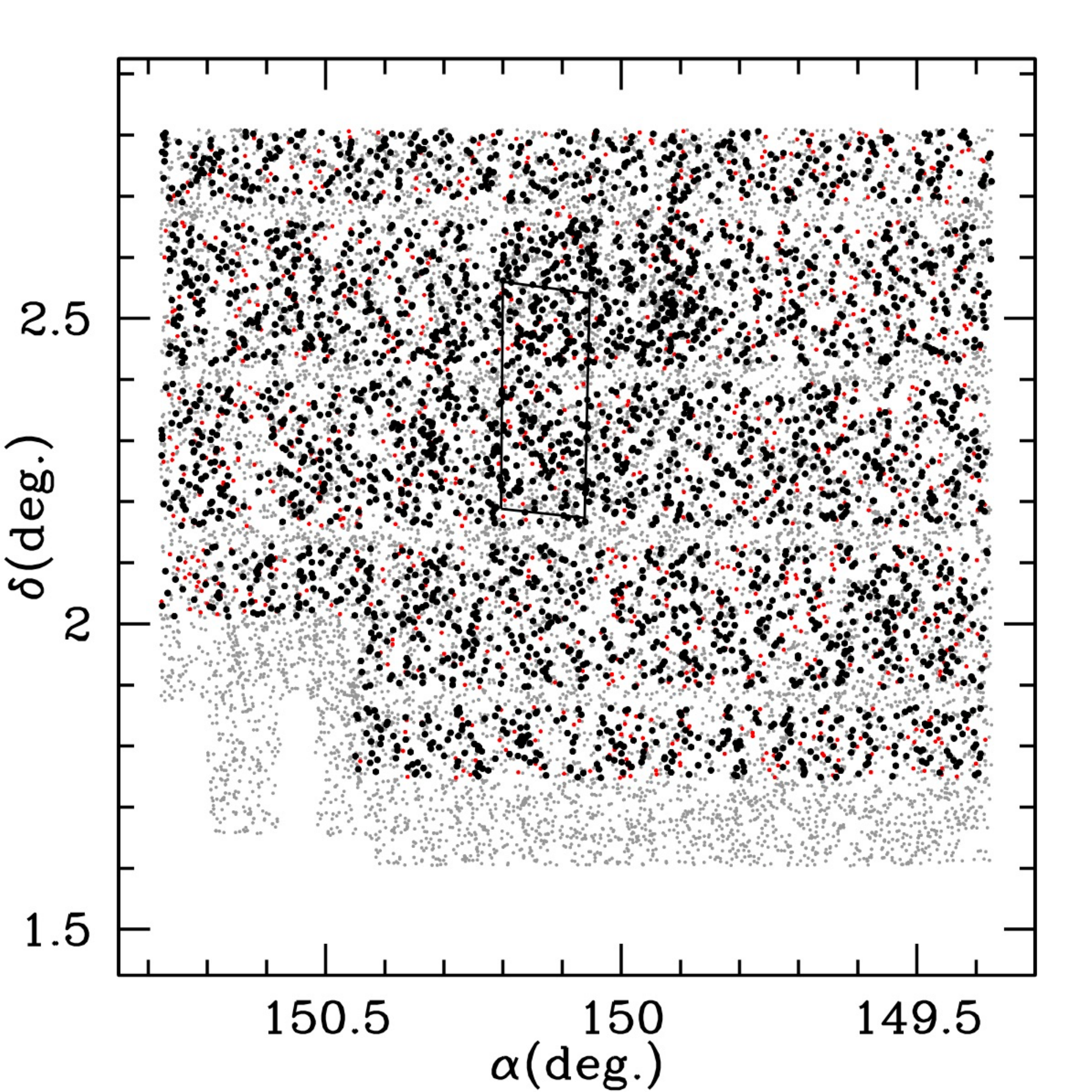}
\caption{{\it Left:} Example of mask design for 1 out of 32
  pointings. The four squares delineate the footprint of the four
  VIMOS detectors.  The dispersion direction is horizontal (the
  R.A.~direction).  The small open circles indicate the full primary
  sample; the larger, filled circles indicate those that have been
  given slits in this mask.  The red symbols are the filler
  galaxies. {\it Right:} Full survey design of 32 pointings, with
  symbols as in the left-hand panel.  Incomplete coverage of
  UltraVISTA leaves the bottom-left corner out of the LEGA-C
  survey. Instead we use those two pointing to re-sample the CANDELS
  field (the rectangular outline near the center) and the well-known
  $z=0.73$ overdensity just to the N-W of the CANDELS field. }
\label{fig:mask}
\end{figure*}

Four masks designs are created for each of the 7.0\arcmin $\times$
8.4\arcmin~VIMOS detectors.  Slits are always in the N-S direction,
1\arcsec~in width and at least 8\arcsec~long.  Slit assigment is fully
automated and proceeds as follows.  First, primary targets more than
100\arcsec~away from the detector edges in the dispersion direction --
to optimize wavelength coverage\footnote{The resulting blue wavelength
  cutoff ranges uniformly from 5800$\AA$ to 6300$\AA$; the red
  wavelength cutoff ranges uniformly from 8300$\AA$ to 9400$\AA$} --
are assigned slits, in order of their K-band flux (brightest first).
This strategy ensures that we maximize the collected number of photons
without biasing the sample toward blue, optically bright galaxies.
This first step of slit assignment typically fills 60\% of the mask.
Before proceeding with assigning additional slits to galaxies, we
include at least one blue star with spectral coverage at the longest
wavelengths to produce a telluric absorption spectrum.  Candidate blue
stars are identified on the basis of their SDSS colors ($g-r < 0.7$).
Next, galaxies in the primary sample closer to the detector edges than
100\arcsec~are included in the masks, increasing the occupation of
mask real estate to $\sim75\%$.  Next, alignment stars are included in
all four masks, followed by galaxies from the filler samples (see
Table 1).  An example of a mask design is shown in Figure
\ref{fig:mask}.  These mask designs are fed into ESO's vvmps software
package to produce the files required for cutting the masks.

The $\sim$2\arcmin~gaps between the detectors and the
100\arcsec~regions near the edges of the detector in the wavelength
(E-W) direction that are preferably avoided when assigning slits leave
about half of the sky uncovered in a single pointing.  However, by
interlacing the pointings in the E-W direction, we achieve 100\%
coverage in that direction, leaving only the $\sim$2\arcmin~detector
gaps in the N-S direction. The survey consists of 32 pointings (four
masks each), covering the 1.6 square degree UltraVISTA footprint, and
including a total of 4,027 galaxies, 3,142 of which are from the
primary sample (see Table 1 for details).

\subsection{Observations}

\begin{table*}
\begin{center}
  \caption{ Summary of past and currently scheduled observations.  Scheduled hours correspond to allocated time; integrated hours refer to the actual on-target exposure time.}
 \begin{tabular}{|l|c|c|l|}
\hline
Run ID & Hrs.~(sched.) & Hrs.~(exp.) & Dates \\
\hline
\hline
\multicolumn{4}{|c|}{Period 94}\\
\hline
194.A-2005(A) & 20 & 13	& 22 Dec 2014 - 27 Dec 2014 \\
194.A-2005(B) & 66 & 39 & 15 Jan 2015 - 26 Jan 2015 \\
194.A-2005(C) & 76 & 41 & 14 Feb 2015 - 24 Feb 2015 \\
194.A-2005(D) & 66 & 17 & 15 Mar 2015 - 26 Mar 2015 \\
\hline
Sum                & 228 & 110 & \\
\hline
\hline
\multicolumn{4}{|c|}{Period 95}\\
\hline
194.A-2005(E) & 50 & 31	& 13 Apr 2015 - 23 Apr 2015 \\
194.A-2005(F) & 30 & 20	& 12 May 2015 - 22 May 2015 \\
\hline
Sum             & 80    & 51 & \\
\hline
\hline
\multicolumn{4}{|c|}{Period 96}\\
\hline
194.A-2005(H) &	26 & 17 & 04 Dec 2015 - 17 Dec 2015 \\ 
194.A-2005(J) &	50 & 32 &  03 Jan 2016 - 15 Jan 2016 \\
194.A-2005(G) &	36 & \nodata & 02 Feb 2016 - 14 Feb 2016 \\
194.A-2005(I) &	72 & \nodata & 02 Mar 2016 - 14 Mar 2016 \\
\hline
Sum             & 184    & \nodata & \\
\hline
\end{tabular}
\label{tab:obs}
\end{center}
\end{table*}

Observations are carried out in visitor mode and in dark time.  Table
\ref{tab:obs} shows past and currently scheduled observing runs in
Periods 94-96.  Similar runs will be scheduled for Periods 97-101,
after which the survey will have been completed barring large delays
due to excessive weather losses or technical problems.

The high-resolution red grating (HR$\_$red) is used in combination
with the GG475 order separation filter.  This provides a typical
wavelength range of $\sim6300{\rm{\AA}}-8800{\rm{\AA}}$ at a
resolution of $R=2500$ and a dispersion of 0.6$\rm{\AA}~$pix$^{-1}$.

For each pointing, we obtain 15-20 observation blocks (OBs) of
4$\times$900s-1200s exposures each; the total integration time per
mask is 20 hours.  Occasionally an OB will be repeated as a result of
degrading weather conditions, and integration times shortened for
scheduling efficiency.  Given the large angular sizes of many of our
targets a dithering scheme would require slits that are
12-20\arcsec~long (proportional to the number of offsets), compared to
the minimum of 8\arcsec~that we adopt in our survey.  In addition, by
using the noisy sky as a background measure we would sacrifice
16\%-30\% in depth (inversely proportional to the number of offsets)
equivalent to losing 6-10 hours of exposure time.  Combining these
factors, dithering would reduce the survey efficiency by a factor of
several.  Fortunately, the low fringing amplitudes of the
red-sensitive detectors that were installed in 2008 allow us to
observe without dithering.  The remaining fringing level and problems
with the subtraction of the brightest atmospheric emission lines for a
subset of targets are merely a cosmetic problem (see Section
\ref{sec:reduc}) and are a far superior choice compared to the cost of
the arguably cosmetically superior dithering approach.

To minimize overheads we deviate from the standard procedure to obtain
calibration data (flats and arcs) immediately after the completion of
an OB, as this would require slewing the telescope to zenith.
Instead, for the data taken between December 2014 to May 2015 we
collected calibration data in batches at a series of rotator angles
that correspond with the rotator angles during the observation of the
science data.  This approach intends to limit residual effects of
changes in flexure and hysteresis to a level that can be appropriately
dealt with during the data reduction process.  Careful analysis of
these calibration datasets then revealed that proximity in time
elapsed between science and calibration data is more relevant than
similarity in rotator angle.  Therefore, we changed our strategy and
obtain a calibration dataset after each block of 2-2.5h science
exposures.

\subsection{Data Reduction}\label{sec:reduc}

The data reduction procedure is split into two parts.  First, we use
the ESO-provided pipeline to produce rectified, wavelength-calibrated
2D spectra.  Second, with a custom-made pipeline we improve the
wavelength calibration, perform simultaneous sky subtraction and
object extraction, and co-add the individual frames.

\subsubsection{ESO pipeline}

The Reflex interface is used to run the ESO-provided pipeline recipes
on the data from an individual OB consisting of four exposures,
producing a single, averaged frame.  The steps performed with the ESO
pipeline include a bias and flatfield correction, slit definition,
slit rectification, wavelength calibration, spectro-photometric
calibration and cosmic ray rejection.

\subsubsection{Custom pipeline}\label{sec:custom}

There are three main reasons for switching to a custom-made set of
data reduction recipes at this point to further process the OB-level,
combined exposures.  First, the VIMOS mask design software forces gaps
between the slits of only 2 pixels (0.4\arcsec), causing bleeding of
bright skylines into the neighboring slits, making sky subtraction
more difficult.  Second, the alignment between the science and
calibration frames is imperfect due to flexure, which requires
reducing the length of the slits across which sky subtraction is
performed.  Third, our non-dithering strategy outlined above requires
that object extraction and sky subtraction are done simultaneously.

2D error spectra are calculated from the flux in the science frames
and the read noise level.  Slit definitions are automatically verified
and adjusted.  Using the expected location of a target galaxy as an
initial estimate, the size and precise location are measured by
fitting a Gaussian profile in the spatial direction after summing
along a 1000 pixel range in the wavelength direction.  This location
is then traced along the entire wavelength range in bins of 100
pixels.  For $\sim$10\% of the targets in each mask there is
insufficient flux to do this, and the location is assumed to be the
expected location and independent of wavelength. Secondary objects in
the slits are extracted separately.

The default model for sky$+$galaxy is a flat background with a
Gaussian profile of a fixed width and location (as measured
above). That is, the two free parameters are the amplitude of the
Gaussian (the galaxy flux) and the sky background level.  This model
is fit to each individual wavelength bin, across the spatial
direction, weighed by the error spectrum.

This simple model does not always produce a sufficiently accurate
description of the data, as made apparent by systematic residuals.  In
such cases a first-order term is added to the background model for
areas where the sky level is high or the gradient in the sky level in
the wavelength direction is large.

This data reduction process produces individual extracted galaxy
spectra and their associated error spectra (for each of the 15-20 OBs,
totalling 20 hours of exposure time). Each of these are corrected for
telluric absorption features as follows: the continuum of the blue
star spectrum is fit by a 5-th order polynomial over the regions that
do not contain telluric absorption features, and divided out.  All
galaxy spectra are then divided by the resulting telluric spectrum.
These are then co-added, weighed by the $S/N$ as measured in the
observed wavelength range $8030{\rm{\AA}} - 8270{\rm{\AA}}$, which is
relatively devoid of bright sky lines. Finally, the flux is measured
over the wavelength range $7000{\rm{\AA}} - 8400{\rm{\AA}}$ and
normalized to match the photometric $i^+$-band flux of the object.
This remedies errors in the automatic spectro-photometric calibration
that can arise due to varying transparency during the observations.
The accuracy of the calibration is better than 5\%, but due varying
slit losses (which range from 10\%-40\%) the precision varies from
galaxy to galaxy.  As a higher-level data product we will provide, in
the future, a new, wavelength-dependent flux calibration based on
matching the shape of the spectrum to that of the full photometric
spectroscopic energy distribution, as well as a model for the slit
losses.

\begin{figure*}[t]
  \includegraphics[angle=0,scale=.63]{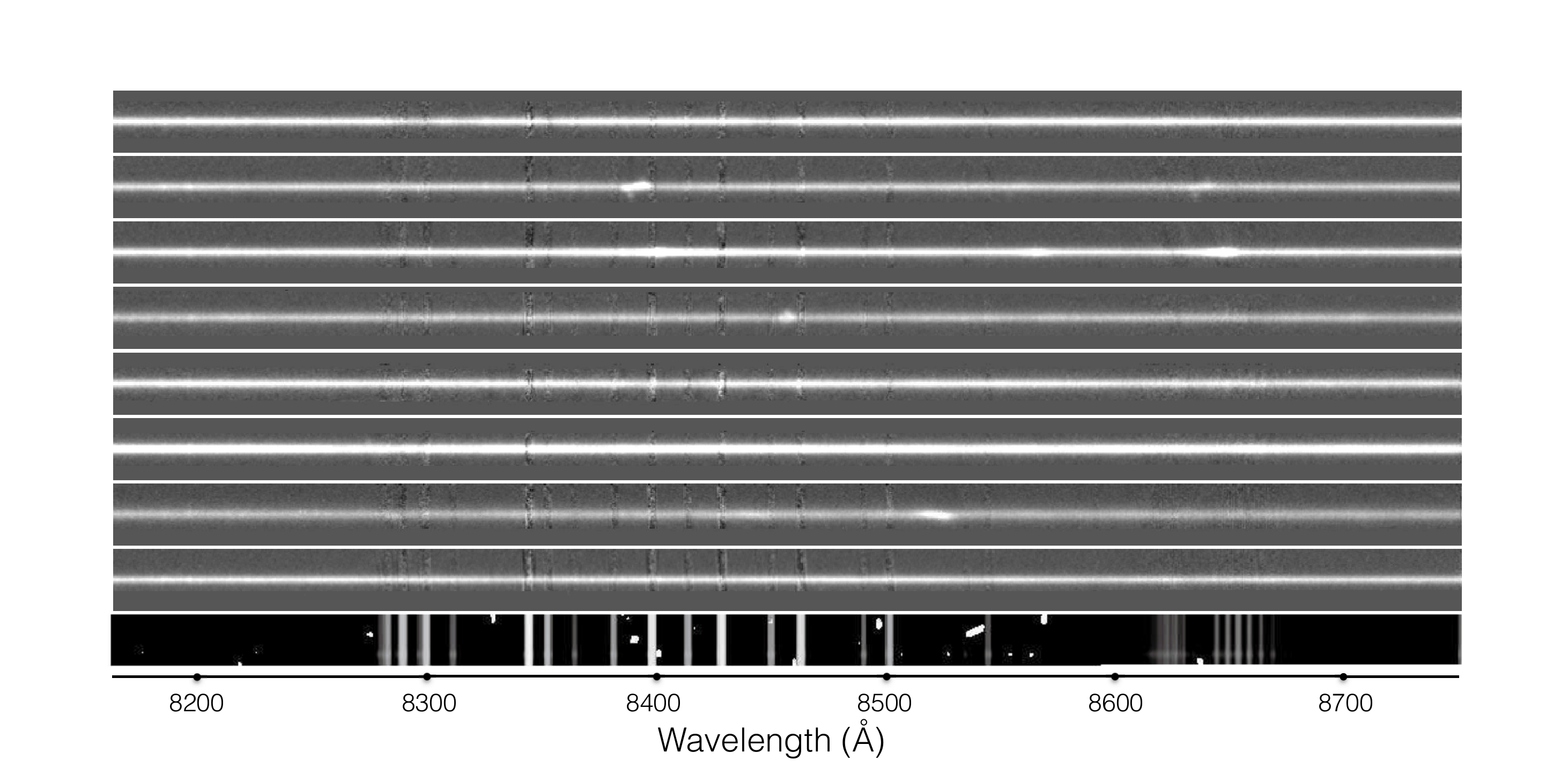}
  \caption{To illustrate the $S/N$ of the data and the quality of the
    sky subtraction we show co-added 2-dimensional spectra of 8
    galaxies over the wavelength range $8260{\rm{\AA}} -
    8760{\rm{\AA}}$, chosen for its bright, telluric emission lines.
    The bottom row shows the variance spectrum of the galaxy spectrum
    right above it, to indicate which parts are clean and which parts
    are affected by the telluric lines. These 8 examples show the full
    range in the quality of sky subraction.  The quality of the sky
    subtraction (see Section \ref{sec:custom} for a description of the
    algorithm) at the locations of the brightest atmospheric features
    varies from galaxy to galaxy due to imperfections in the
    calibration data (bottom two spectra) or low-level fringing (3rd
    from the top).  In other cases the sky subtraction is noise
    limited.  We note that even for the worst cases only a few dozen
    pixels across the entire spectrum are {\it not} noise limited.}
\label{fig:2dspec}
\end{figure*}

As mentioned above, our observing strategy is aimed at maximizing the
survey efficiency and depth by not dithering, a choice that comes at
the cost of retaining some cosmetic flaws as a result of low-level
fringing and imperfections in the sky subtraction of the brightest
atmospheric emission lines that occur when data are obtained under
sub-optimal weather conditions and for objects that are near the edges
of the detectors.  The resulting features are illustrated in Figure
\ref{fig:2dspec}.  We stress that, in total, only about 2-3\% of the
pixels suffer from such problems, and those are wavelength regions
where the night sky is bright anyway, implying that only low-$S/N$
data is lost.

\subsection{Spectral Fitting -- Redshifts, Stellar Population Properties and Velocity Dispersions}\label{sec:fullspec}

Redshifts are measured by cross-correlating with a small set of
general template spectra.  In 98\% of the cases our redshifts are
within the targeted range of $z=0.6-1.0$ and vidually verified.  Among
the 2\% of galaxies that fall outside this range (18 objects), nine
are only marginally lower than $z=0.6$ or higher than $z=1.0$ ($\Delta
z\lesssim 0.02$), and we do not consider their inclusion a mistake.
Among the nine others, four had wrong previous spectroscopic redshifts
(while the photometric redshifts were indeed correct), presumably due
to the poor quality of that previously available spectroscopic data.
Four other galaxies, with no pre-existing spectroscopic redshifts, had
wrong photometric redshifts: 3 of those were edge-on (dusty) disks.
One galaxy -- a blend of two galaxies at different redshifts -- had a
wrong photometric redshift as well as a wrong pre-existing
spectroscopic redshift.  We conclude that our sample is affected by
wrong selection redshifts at the 1\% level.

The key measurements made from the spectra are
the stellar population age and metallicity, and the stellar/gas
velocity dispersion.  Here we provide a brief overview of the
methodology, while a full description of the testing and error budget
are postponed until these measurements are made publicly available
(end of 2016).

The first step is to fit the full spectrum following the method of
\citet{pacifici12}.  Star formation- and chemical enrichment histories
from the \citet{de-lucia07b}~galaxy formation model are used to
generate a library of template spectra, where the evolving chemical
enrichment is calculated based upon the star formation history.  We
estimate gas-phase metallicities as well and CLOUDY \citep{ferland98}
is used to calculate the contribution of nebular line and continuum
emission in a manner consistent with the radiation output of the
stars.  We adopt the model of \citet{charlot00} to treat dust, where
the light from old stars is attenuated by a large-scale, diffuse
component, and the light from young stars is attenuated first by a
birth-cloud component and then by the diffuse component.

Given the restframe spectrum from which the continuum has been
removed, likelihoods are computed for all models, resulting in
marginalized probability distributions for the model parameters.  The
quantities of interest here are specific star formation rate,
luminosity- and mass-weighted age, gas-phase and stellar metallicity.

The ten best-fitting model spectra from this full spectral fit of a
given galaxy spectrum are used as templates for the velocity
dispersion measurement for that galaxy, following the approach of
\citet{bezanson15}. Using the pPXF code \citep{cappellari04},
emission-line subtracted galaxy and model spectra are continuum
filtered by multiplicative and additive polynomials, after which a
Gaussian convolution of the model spectra in real space is fit to the
galaxy spectrum.  The gas-phase velocity dispersion is estimated by
fitting the continuum subtracted galaxy spectrum with the best-fitting
emission line model spectrum, convolved with a Gaussian.  We note that
these kinematic quantities are integral measurements from the light in
the slits -- no aperature corrections are made at this point.

The resulting stellar velocity dispersion is then used to smooth the
model spectra used for the spectral fitting and the initial stellar
population fit is repeated to verify that the initial results were not
biased due to the mismatch in spectral resolution.

\subsubsection{Stellar Populations: Line Indices}

Measurement and interpretation of stellar absorption features in the
optical range provide a well-tested and alternative method to full
spectral fitting to derive constraints on stellar mass, ages and
abundances. In particular we measure all the indices defined in the
Lick/IDS system \citep{worthey94, worthey97}, including the high-order
Balmer lines, in the galaxy spectra cleaned of emission lines
(obtained by subtracting the best-fit emission line model spectrum as
described in Section \ref{sec:fullspec}).  Following the approach
outlined in \citet{gallazzi05, gallazzi14}, we derive marginalized
probability density functions for stellar mass, light- and
mass-weighted stellar ages and stellar metallicities by comparing the
observed strengths (accounting for velocity dispersion broadening) of
optimal sets of absorption indices (defined on the basis of the galaxy
redshift) to a library of BC03-based models convolved with stochastic
SFHs and metallicities.  Additionally, element abundance ratios are
estimated from the ratio of Mg and Fe features \citep[e.g.,~$\rm
  Mgb/\langle Fe\rangle$]{gallazzi06} and from CN indices
\citep[e.g.,][]{carretero04}.

\subsection{Auxiliary Datasets}

\subsubsection{Photometry: Stellar Masses and Rest-frame Colors}
Using the data from \citet{muzzin13a,muzzin13} and following their
methodology we use FAST \citep{kriek08} to fit the photometric
spectral energy distribution and measure stellar masses and rest-frame
colors.  Here, a template library with solar-metallicity spectra from
SFHs with exponentially declining SFRs and dust-screen attenuation
\citep{calzetti00} is used.  The improvements with respect to the
modeling results from \citet{muzzin13} stem solely from the use of our
spectroscopic redshifts.  In a second iteration we jointly fit the
photometry and spectroscopy, deriving stellar masses and rest-frame
colors with the methodology described above in Section
\ref{sec:fullspec}.  We will provide both photometric and
photometric$+$spectroscopic stellar mass estimates.

\begin{figure}
  \includegraphics[scale=.43]{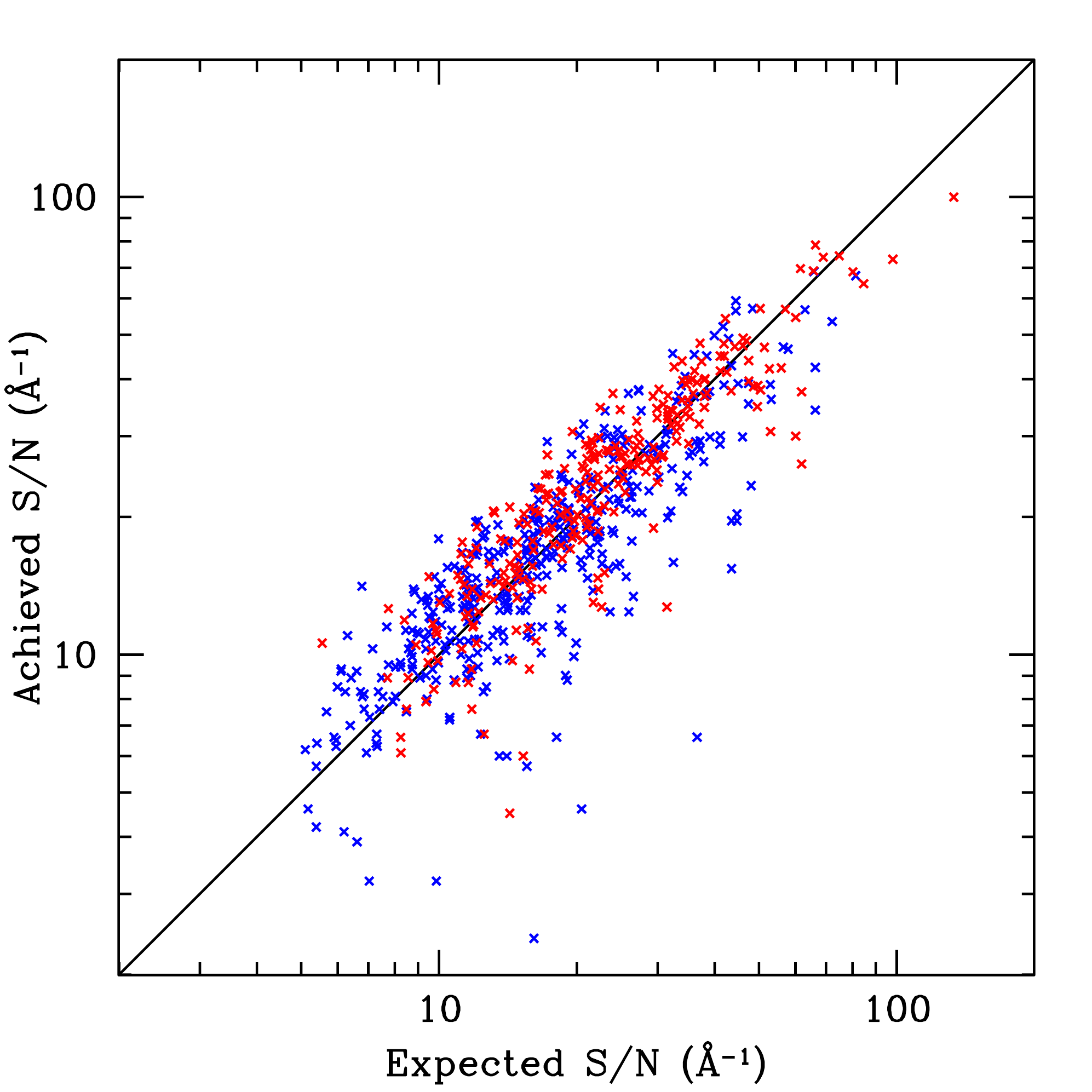}
\caption{Measured $S/N$ per $\rm{\AA}$ (averaged over $8030\rm{\AA} -
  8270 \rm{\AA}$) for the first year of data versus the expected
  $S/N$, based on $i^{+}$ band magnitude and galaxy size, calculated
  in preparation of the survey.  Blue crosses indicate star-forming
  galaxies, red crosses quiescent galaxies.  The general agreement
  implies that our predictions were accurate and that the data quality
  is as expected, if not slightly better.  The scatter can be
  explained by variations in radial surface brightness profiles, axis
  ratios, position angles, and observing conditions.}
\label{fig:sn}
\end{figure}
  
\begin{figure}
\includegraphics[scale=.43]{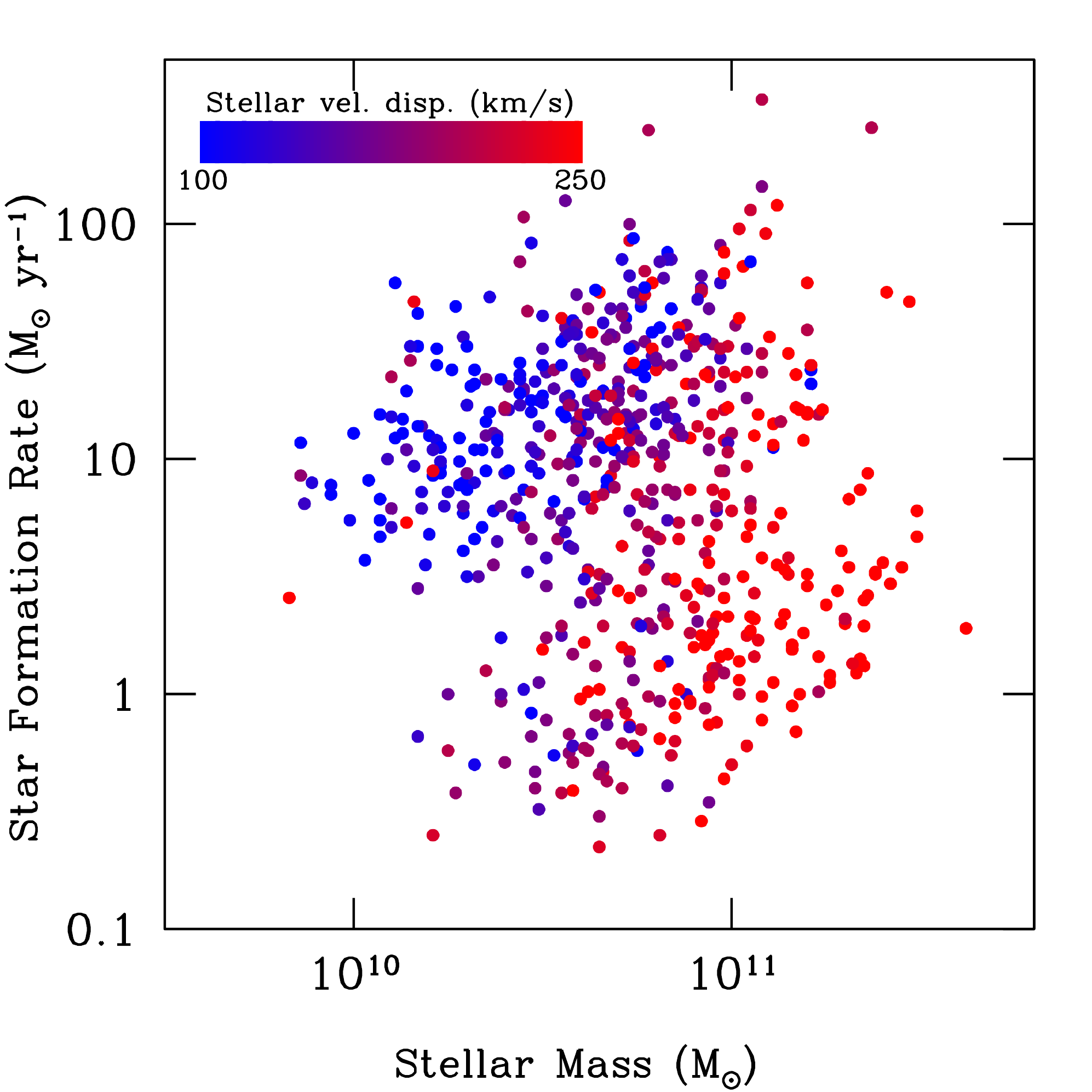}
\caption{Stellar mass vs.~UV$+$IR-based SFR for the LEGA-C 1st-year
  primary sample (653 galaxies).  The color-coding represents stellar
  velocity dispersion.}
\label{fig:m_sfr}
\end{figure}

\subsubsection{Structural Parameters from HST imaging}
ACS imaging is available for nearly all targets from the COSMOS
program \citep{scoville07a}.  10\arcsec~image cutouts are created with
the IPAC tool and the 2-D light profiles are fit with single-S\'ersic
profiles using {\tt galfit} \citep{peng10}, with effective radius,
total magnitude, S\'ersic index, axis ratio, and position angle, and
background as the free parameters.  The Point Spread Function is
chosen from a list of stars in the COSMOS field based on closest
proximity in terms of detector position.  A user-created error image
is calculated based on the IPAC-provided weight image with added
Poisson noise from the objects.  Neighbors are simultaneously fit.
Uncertainties are assigned based on the total $S/N$ and values of the
fitting parameters, as described by \citet{van-der-wel12}.

\section{First Look Results}\label{sec:firstlook}

An early milestone in the LEGA-C survey is to confirm that the data
quality is as expected.  In Figure \ref{fig:sn} we show the $S/N$
distribution for the primary sample observed in 2014-2015, in
comparison with the expected $S/N$ that we calculated in advance.  For
star-forming galaxies we typically achieve
$S/N=10-20\rm{\AA}^{-1}$~and for quiescent galaxies
$S/N=20-30\rm{\AA}^1$, both in line with the expected values.

In Figure \ref{fig:m_sfr} we show for the primary sample observed so
far (23\% of the total) the bi-modal distribution in terms of stellar
mass and star-formation rate \citep[broad-band photometry-based,
  from][]{muzzin13}\footnote{We note that the star-formation rate
  indicator is based on UV and IR fluxes.  Therefore, any source of
  UV+IR radiation that is not from star formation is still counted as
  such.  For example, dust heated by evolved stellar populations may
  significantly affect the IR luminosity of massive early-type
  galaxies \citep[e.g.,][]{groves12}.} The dynamic range in these
quantities is such that our sample represents the full galaxy
population with stellar masses $M_* > 10^{10}~\msol$, where the
majority of stars form and live throughout cosmic history
\citep[e.g.,][]{karim11}.

An illustration of the application of new information obtained through
our deep spectroscopy is shown in Figure \ref{fig:m_sfr}, where we
color-code the objects by their stellar velocity dispersion.  Galaxies
with higher stellar masses have higher velocity dispersions, but also,
at a fixed stellar mass quiscent galaxies have higher velocity
dispersions than star-forming galaxies. This has been seen for
present-day galaxies \citep{wake12} and has been predicted for
high-redshift galaxies \citep{franx08}.  Our evidence that star
formation activity correlates with dynamical structure implies that
the different size-mass relations for star-forming and quiescent
galaxies \citep[e.g.,][]{shen03, van-der-wel14} provide important
constraints on galaxy formation models, and do not merely reflect
differences in $M/L$ gradients and not result from practical problems
with measuring sizes of distant galaxies.

We note that the correlation between true velocity dispersion (meaning
the dispersion in stellar velocities at a fixed location in a galaxy)
and star-formation activity must be even stronger than is apparent in
Figure \ref{fig:m_sfr}: much of our velocity dispersion estimates --
simple, 2nd-order velocity moments, measured from the integrated light
in the slit -- presumably reflect rotation, especially for
star-forming disks.

\begin{figure*}[t]
\includegraphics[angle=90,scale=0.65,trim={3cm 0cm 2cm 1cm},clip]{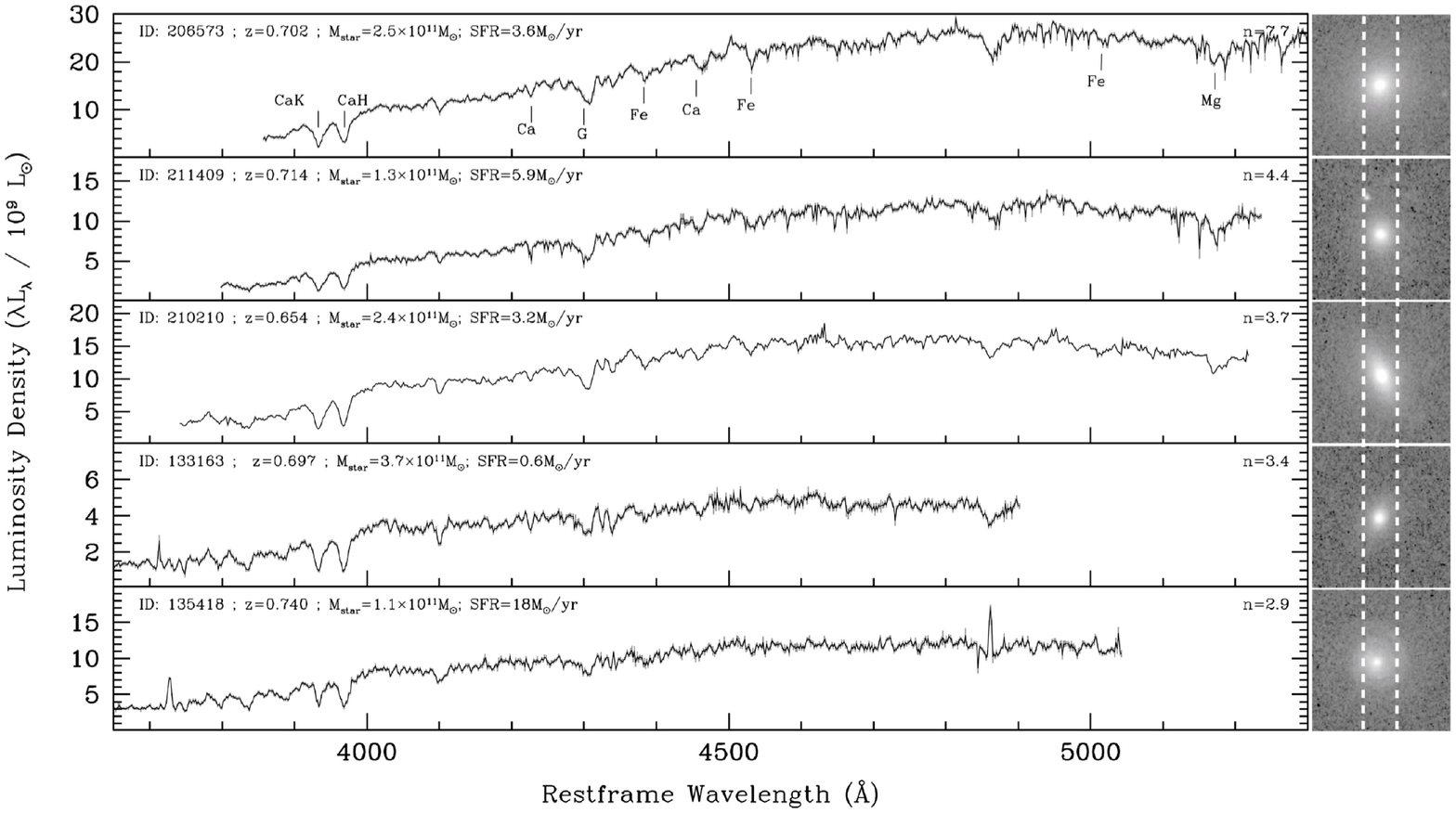}
\includegraphics[angle=90,scale=0.65,trim={3cm 0cm 2cm 1cm},clip]{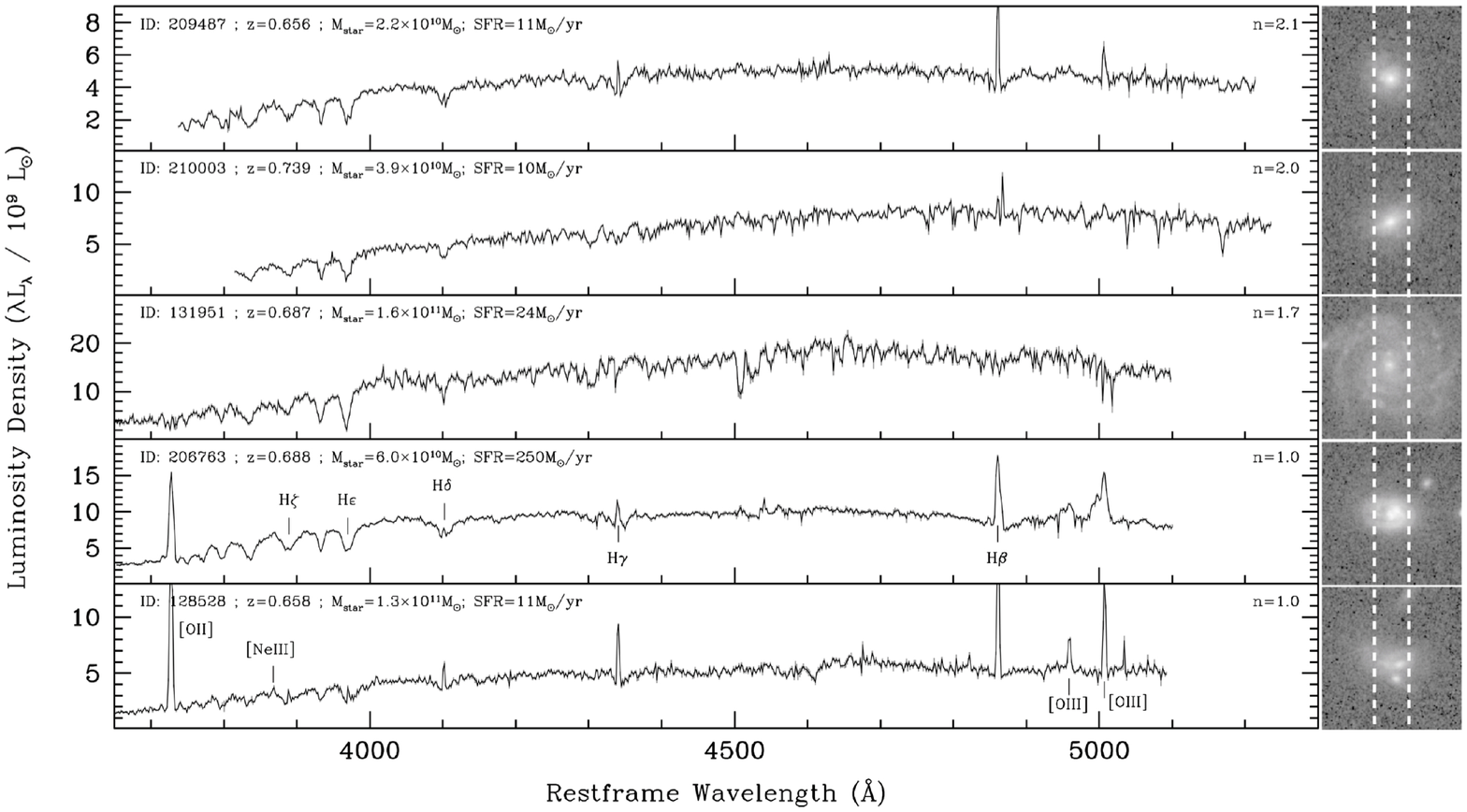}
\caption{Ten representative LEGA-C spectra sorted by Sersic index $n$.
  These spectra have $S/N$ close to or better than the survey median
  -- 1$\sigma$ errors are indicated in grey.  Metal absorption
  features are labeled in the top panel; Balmer absorption/emission
  lines and metal emission lines are labeled in the bottom panels.
  Identification number, redshift, stellar mass (from broad-band SED
  fits), and SFR (derived from total UV and mid-IR fluxes) are given
  in each panel.  On the right-hand side HST/ACS cutouts (4.5\arcsec
  across) are shown, along with the slit orientation (N-S) and width
  (1\arcsec).  A clear correlation between spectral and structural
  properties is obvious at first sight: more concentrated galaxies
  have older populations than less concentrated galaxies. In many
  cases slit losses are limited and the spectra provide a
  representative view of the galaxy, but in some cases a significant
  amount of the light is missed due to the extent and/or orientation
  of the galaxy (e.g., ID 131951).}
\label{fig:spec_sersic}
\end{figure*}

A visual impression of the quality of the spectra, and the range in
galaxy properties in the same, is shown in Figure
\ref{fig:spec_sersic}.  Ten spectra are sorted by the galaxies'~Sersic
index, a parameter that unlike stellar mass or color is measured
independently of spectral properties.  Stellar population
characteristics are visually apparent for all ten galaxies, regardless
of their structural properties or star formation activity.  We see a
clear change in spectral properties as a function of structure, from
passive/old for high-Sersic index galaxies to young/star-forming for
low-Sersic index galaxies.  We note that even the youngest objects
show clear absorption lines.

\begin{figure*}[t]
\includegraphics[angle=90,scale=0.65,trim={3cm 0cm 2cm 1cm}]{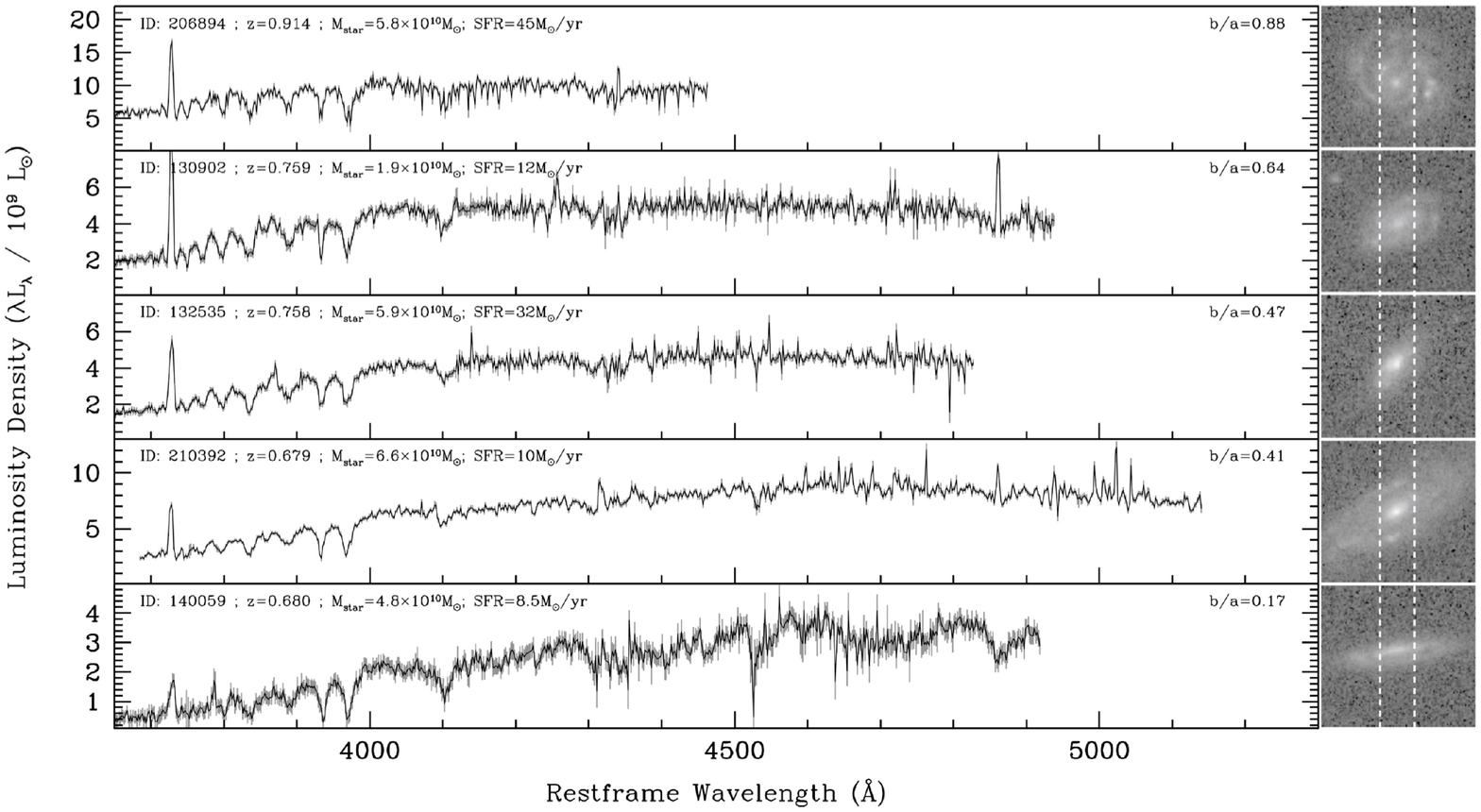}
\caption{Five LEGA-C spectra of disk galaxies sorted by inclination
  (projected axis ratio $b/a$).  See caption of Figure
  \ref{fig:spec_sersic} for a detailed description of the contents of
  this figure.  A clear reddening is seen for more inclined galaxies,
  and despite significant dust columns continuum is with high $S/N$ in
  all cases (e.g., ID 140059).}
\label{fig:spec_incl}
\end{figure*}

A further illustration is given in Figure \ref{fig:spec_incl}, where
we sort five star-forming disk galaxies by their inclination angle.
Face-on and edge-on disks each present their own specific
observational challenge for deep spectroscopy: low surface brightness
and extinction, respectively.  Yet, we have achieved high-$S/N$
continuum spectra for each of the galaxies.

\section{Timeline and Outlook}

The LEGA-C observations will nominally conclude in May 2018, modulo
uncertain weather and other losses. The first data release is
scheduled for June 1st, 2016. This release will include reduced,
calibrated, extracted spectra for all 925 galaxies for which
observations were completed before June 2015.  A basic catalog with
UltraVISTA IDs, coordinates, magnitudes, redshifts, and $S/N$ values
will accompany this release.

A second release will follow by the end of 2016, which will include
spectra for all galaxies observed until May 2016, as well as the first
set of value-added catalogs with redshifts, velocity dispersions, and
emission/absorption line strength indices.  The third and fourth data
releases are scheduled for December 2017 and December 2018.  Besides
the additional spectra and first set of value-added catalogs, these
two releases will also include a second set of value-added catalogs
containing stellar masses, ages, metallicities, structural parameters
from HST imaging and dynamical masses.

With the data in hand, we can, as a community, start addressing
scientific questions outlined in Section 2.  The data will be publicly
released by June 1st 2016, and we do not have to wait until the sample
collection is complete to achieve many science goals.  The current
dataset already far surpasses any previously existing dataset.

We thank the referee for thorough reading of the manuscript and their
constructive comments.  AW and KN acknowledge support from the
Deutsche Forschungsemeinschaft (GZ: WE 4755/4-1).  RB gratefully
acknowledges support by NASA through Hubble Fellowship grants
\#HF-51318 awarded by the Space Telescope Science Institute, which is
operated by the Association of Universities for Research in Astronomy,
Inc., for NASA, under contract NAS 5-26555.  CP acknowledges support
by an appointment to the NASA Postdoctoral Program at the Goddard
Space Flight Center, administered by USRA through a contract with
NASA.  We gratfeully acknowledge the NWO Spinoza grant.  VW
acknowledges support from the European Research Council Starting Grant
(SEDMorph; P.I. V.~Wild).

\bibliographystyle{../bibtex/apj.bst}

\end{document}